# On the junction physics of Schottky contact of (10, 10) MX$_2$ (MoS$_2$, WS$_2$) nanotube and (10, 10) carbon nanotube (CNT): an atomistic study

Amretashis Sengupta[1,2*]

[1]*Hanse-Wissenschaftskolleg (HWK), Lehmkuhlenbusch 4, 27753 Delmenhorst, Germany*
[2]*Bremen Center for Computational Materials Science (BCCMS), Universität Bremen, Am Fallturm 1, 28359 Bremen, Germany*

*Corresponding Author: e-mail:* sengupta@bccms.uni-bremen.de

**Abstract:** *Armchair nanotubes of MoS$_2$ and WS$_2$ offer a sizeable band gap, with the advantage of a one dimensional (1D) electronic material, but free from edge roughness and thermodynamic instability of nanoribbons. Use of such semiconducting MX$_2$ (MoS$_2$, WS$_2$) armchair nanotubes (NTs) in conjunction with metallic carbon nanotubes (CNT) can be useful for nanoelectronics and photonics applications. In this work atomistic simulations of MoS$_2$ NT-CNT and WS$_2$ NT-CNT junctions are carried out to study the physics of such junctions. With density functional theory (DFT) we study the carrier density distribution, effective potential, electron difference density, electron localization function, electrostatic difference potential and projected local density of states of such MX$_2$ NT-CNT 1D junctions. Thereafter the conductance of such a junction under moderate bias is studied with non-equilibrium Green's function (NEGF) method. From the forward bias characteristics simulated from NEGF, we extract diode parameters of the junction. The electrostatic simulations from DFT show the formation of an inhomogeneous Schottky barrier with a tendency towards charge transfer from metal and chalcogen atoms towards the C atoms. For low bias conditions, the ideality factor was calculated to be 1.1322 for MoS$_2$ NT-CNT junction and 1.2526 for the WS$_2$ NT-CNT junction. The Schottky barrier heights displayed significant bias dependent modulation and are calculated to be in the range 0.697 - 0.664eV for MoS$_2$ NT-CNT and 0.669-0.610 eV for the WS$_2$ NT-CNT respectively.*

## 1. INTRODUCTION

1D material like nanotubes are particularly interesting in nanoelectronics as they offer enhanced thermodynamic and chemical stability compared to 2D materials owing to the absence of edge-effects. [1]-[3] Over the years CNTs have been extensively studied for their nanoelectronics applications in MOSFETs and as future interconnect materials in microchips. [4]-[5] In the present decade transition metal dichalcogenides (general formula MX$_2$, M=transition metal, X=chalcogen) have emerged as a very promising class of layered crystals suited for nanoelectronics owing to their non-zero band gap.[6]-[8] In this regard, nanotubes of MoS$_2$ and WS$_2$ have also been experimentally reported [9]-[13], and are beginning to be explored for their potential use as electronic material. The feasibility of such MoS$_2$, WS$_2$ nanotube (NT) based electron device has been shown in a recent study by Levi et. al. [13]. Also recently composites made up of MX$_2$ NT and CNT have shown good promise for application in Li-ion batteries. [14] The MoS$_2$ and WS$_2$ armchair NTs (hereafter referred to commonly as MX$_2$ NT), which offer band gaps of 1.5 – 2 eV, [15]-[17] may be combined with the metallic/ narrow band-gap CNTs for nanoelectronic in future nanoscale devices and interconnects. Based on the successful experimental realization of MX$_2$ NT FET and CNT FET, [4, 5,13] the prospect of a MX$_2$ NT-CNT junction formed with nano-manipulation (of individual CNT and MX$_2$ NT) or other advanced VLSI fabrication techniques is quite high. Due to the complimentary optical absorption properties of CNT and MX$_2$ NT, such junction would have potential applications in photovoltaic and optoelectronic devices as well. Thus an in-depth computational study of MX$_2$ NT-CNT junctions warrants merit.

Two MX$_2$ armchair NTs (namely MoS$_2$ and WS$_2$) were chosen due to their successful experimental realization reported in literature. [9]-[13] With density functional theory (DFT) we simulate the electrostatic features of the junction such as effective potential and electron density, electron difference density (EDD), electron localization function (ELF). We also simulate the electrostatic difference potential (EDP) and projected local density of states (PLDOS) of the CNT-MX$_2$ junction. The output characteristics of such a junction under moderate bias is studied with non-equilibrium Green's function (NEGF) method. From the simulations we further extract diode parameters such as Schottky barrier height, ideality factor, diode series resistance of such junctions with the activation energy method [18]-[21] and the Cheung method [20].

## 2. METHODOLOGY

In our supercell (Fig. 1) we have a (10,10) MoS$_2$ / WS$_2$ armchair NT forming a junction with a (10,10) armchair CNT. (10,10) CNT is chosen for its metallic property. The (10,10) MX$_2$ nanotube with an outer diameter of ~ 20.5 $\text{Å}$ is chosen for our studies as MX$_2$ NTs band gaps widen with increasing diameter (due to dissimilar strain on inner and outer layers of chalcogen atoms). [22,23] In this way we try to achieve a moderate band gap offset between the two materials. Density functional theory (DFT) calculations has been carried out in AtomistixToolKit 2015.1 [24] with the Perdew-Zunger (PZ) exchange correlation within the local density approximation (LDA). [25, 26] The pseudopotential sets used are Troullier-Martins type norm-conserving double Zeta polarized (DZP) basis sets [24, 27] (FHI [Z=6] DZP for Mo, W and S and FHI [Z=4] DZP for C). We employ a 1x1x9 Monkhorst-Pack [28] k-point grid with the cutoff energy of 100 Ry. The Pulay mixer algorithm [29] controls the self-consistent iterations with 0.0002 Ry tolerance and 100 maximum steps. The structures are relaxed to a maximum force of 0.05 $eV/Å$ and maximum stress (for cell optimization) of 0.05 $eV/Å^3$ with a limited memory Broyden-Fletcher-Goldfarb-Shanno (LBFGS) algorithm. [30]

The electron difference density (EDD) $\rho_{ED}$ is basically the difference between the superposed atomic valence density and self-consistent charge density.[24] The electrostatic difference potential which is approximately equal to change in Hartree potential ($\Delta V_H$), is obtained by solving the Poisson's equation considering $\rho_{ED}$ as the charge density. [24]

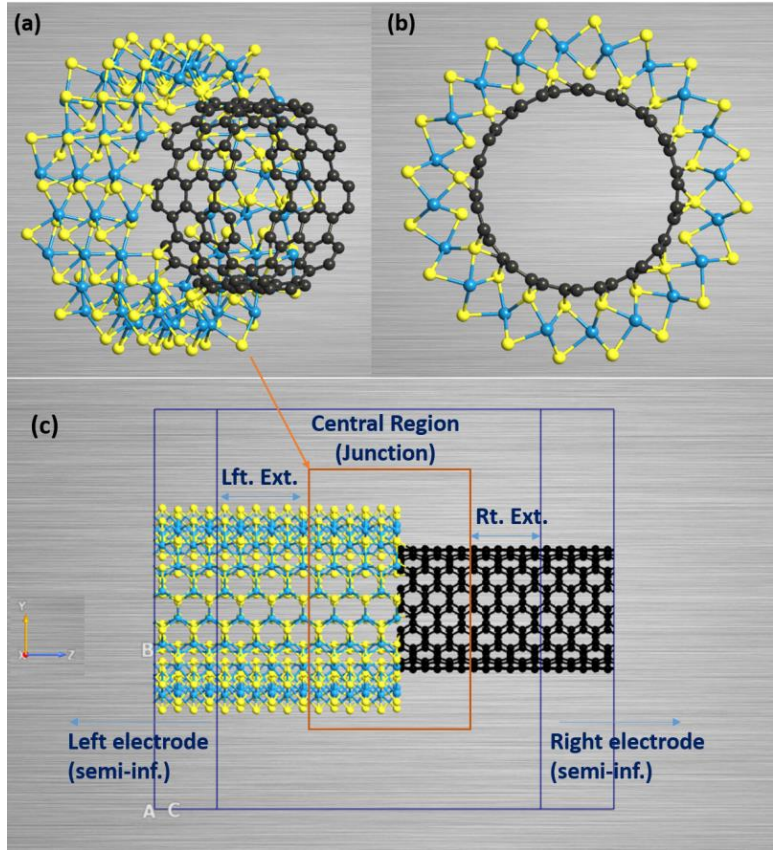

**Fig. 1:** (a) Off axis and (b) axial view of the MX$_2$ NT-CNT junction (c) the device supercell used in calculations.

For DFT-NEGF simulations we consider the junction as a two probe system (Fig.1) with MX$_2$ NT on one side and CNT on other side being extended as perfect (ideal) electrodes. The device supercell (shown in Fig. 1(c)) is a combination of periodic and finite configurations. The central region where the interface between the two different materials physically exists, shown in Fig. 1(a) is basically a finite (non-periodic) region. Also there exists regions known as the electrode extensions (labeled Lft. Ext. and Rt. Ext. in Fig. 1(c)) in both ends of the central region, which are essentially identical in composition and lattice matched with the respective electrodes. [24] The periodic and semi-infinite electrodes are constructed by internal routines of the QuantumWise ATK tool which recognize the principal layers from the left/right side of the central region and repeat them accordingly.[24] The electrodes in ATK are modeled as perfect leads. Sufficient vacuum is maintained on directions orthogonal to the transport for allowing the decay of electron density and electrostatic potentials and to avoid spurious image interactions (from unwanted repetitions). Also since it is the junction which actually provides the functionality to the device, and is our point of interest, all the plots presented are for this central region (junction).

For the simulations, we use 1x1x100 k-points to sample the device supercell with a 2 D fast Fourier transform (FFT2D) Poisson solver. [24] On the electrode faces of the device supercell, the Dirichlet boundary conditions are applied and on all the other faces the boundary condition is set to periodic. The average Fermi level is taken as the energy zero parameter for the calculations with Krylov self-energy calculator. To find the ideality factor of the junction we follow the method used by Stradi et. al. [18] to fit the forward bias characteristics to the relation [18]-[21]

$$I\left(1-e^{-\beta V_{Bias}}\right)^{-1} = I_0 e^{\frac{\beta V_{Bias}}{\eta}} \tag{1}$$

$I_0$ is the saturation current, $\beta = \frac{q}{k_B T}$ where $q$ is electronic charge, $k_B$ the Boltzmann constant and $T$ temperature. $V_{Bias}$ is the applied bias voltage and $\eta$ is the ideality factor. The Schottky barrier height from the activation energy method $\Phi^{AE}$ is calculated from the current-temperature $(I-T)$ characteristics by fitting to the formula [18]

$$\frac{d\left[\ln(I/T^2)\right]}{d\left[1/T\right]} = \frac{q}{k_B}\left(\frac{V_{Bias}}{\eta} - \Phi^{AE}\right) \tag{2}$$

From the knowledge of $\eta$, $\Phi^{AE}$ the junction resistance defined as $R^{jn} = RA_{eff}$ is calculated from the thermionic emission formula using Cheung's method. [20,21]

### 3. RESULTS & DISCUSSION

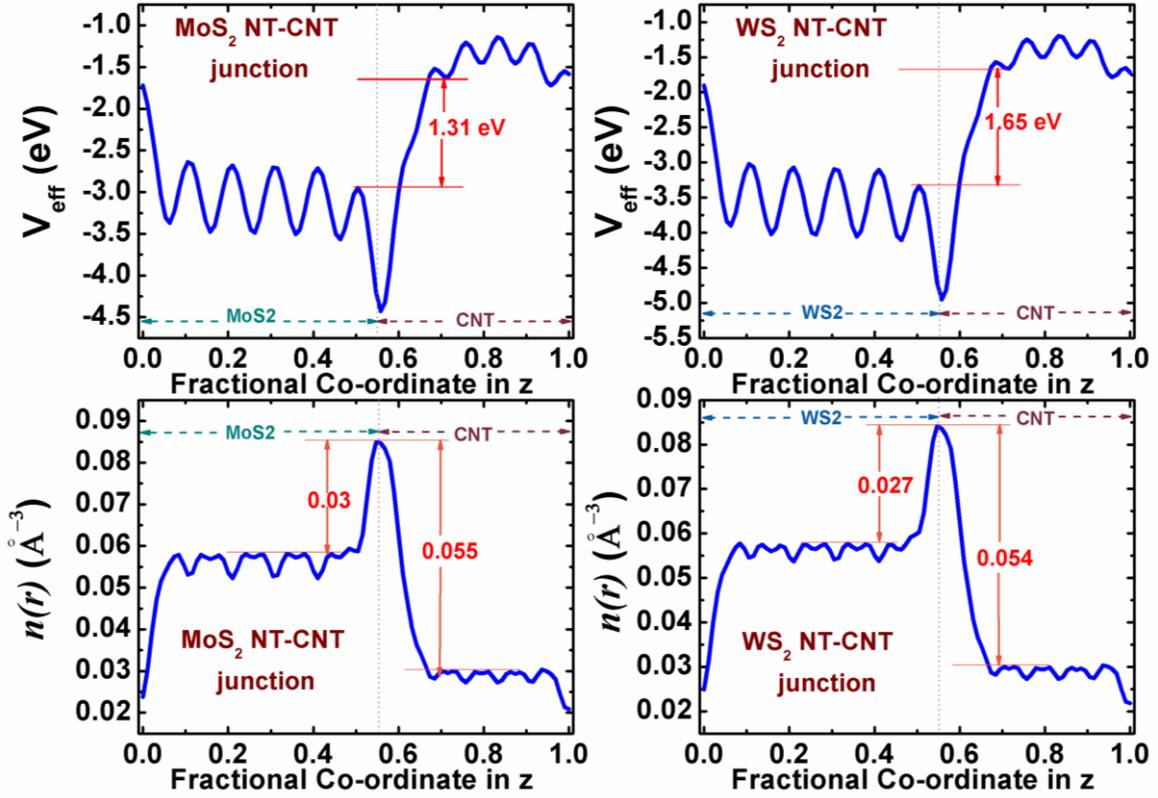

**Fig. 2:** The projected effective potential (top) and the projected carrier density (bottom) plots along the z-axis of the $MX_2$ NT-CNT junction.

In Fig. 2, we have the projected effective potential $V_{eff}$ and the projected carrier density plots along the c-axis (z-axis) of the $MX_2$ NT-CNT junction (central region). From the effective potential plots, a $V_{eff}$ difference of ~1.31 eV for $MoS_2$ NT-CNT and 1.65 eV for $WS_2$ NT-CNT junction is observed. Also at the junction, a significant amount of charge accumulation at the interface is seen in the electron density plots. The electron distribution profiles have very similar bearings and magnitude for both $MoS_2$ and $WS_2$ junctions. The charge density does not seem to die down very quickly into the CNT region suggesting a possible screening effect at the junction.

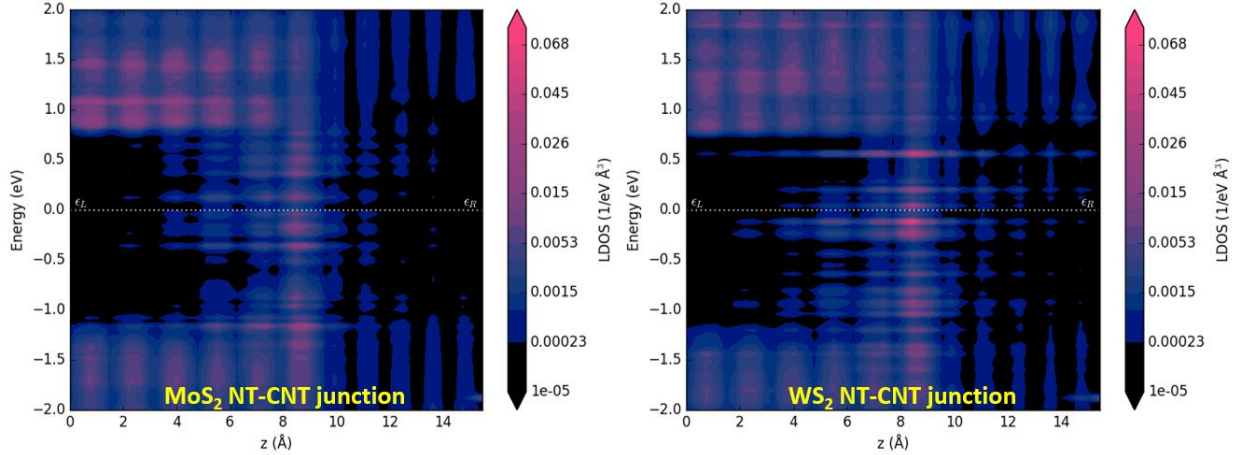

**Fig. 3:** Projected local density of states (PLDOS) of the $MX_2$ NT-CNT junctions

The projected local density of states (PLDOS) at the junction under equilibrium shows the formation of a large number of metal induced gap states (MIGS) in the semiconducting side of the junction. The PLDOS is seen to be significantly increased at the CNT-$MX_2$ interface. The unconventional nature of the junction with two 1 D materials and the unique geometry of the interface gives rise to a PLDOS much different from that of conventional metal-semiconductor junctions. The scattered puddles of metallic states on the CNT side is something different than conventional metals and the periodic zones of high and low LDOS on both sides is a result of the spatial distribution of varying types of atoms (metals, sulfur and carbon). Overall the depletion width of the $WS_2$ NT-CNT junction seems to be somewhat thin in dimension but having higher LDOS as compared to that of the $MoS_2$ NT-CNT junction. Also the MIGS tend to extend deeper into the $MoS_2$ side as compared to that in the $WS_2$. The PLDOS overall suggest a rather inhomogeneous nature of the Schottky junction thus formed between these two kinds of nanotubes.

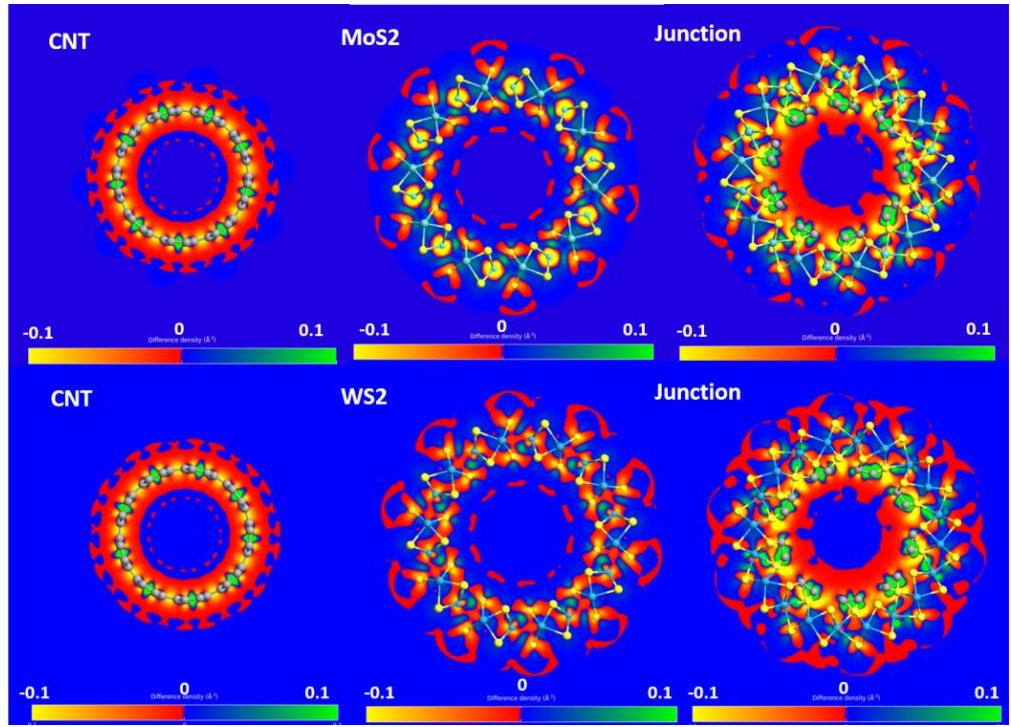

**Fig. 4:** The electron difference density (EDD) plots for the MX$_2$ NT-CNT junctions in the *xy* cut plane.

Fig. 4 shows the electron difference density (EDD) plots for the MX$_2$ NT-CNT junctions. The cut-plane is the *xy* plane at different z co-ordinates to show the EDD in the different regions of the supercell namely the CNT side, MX$_2$ side and at the physical interface (junction). EDD is basically the difference between the superposed atomic valence density and self-consistent charge density. [24] A transfer of charge from one atom to another, results in a positive EDD on the charge acceptor and a negative EDD on the donor. We use a step color bar where Blue-Green indicates positive EDD and Red-Yellow indicates negative EDD. At the MoS$_2$ NT-CNT junction if we look at the EDD on the S atoms and Mo atoms (lying in-plane) the EDD has a negative sign. As we move towards an S atom along the Mo-S bond, near the midway point there is a small region of positive EDD, but moving further closer to the S atom the EDD reverts back to negative. The negative EDD of the inner S atoms are more pronounced than that of the outer S atoms. If we see the S-C bonds at the junction the region near C atoms, (including those around the S-C and C-C bonds) shows a positive EDD. These results indicate a tendency to transfer electrons from the Mo and S atoms towards the C atoms at the junction. This is also inferred from the electron density plots in Fig. 2. In case of the WS$_2$ NT-CNT junction, the EDD plots are slightly different as the W atoms in particular do not show much significant predisposition towards the positive/negative EDD as the Mo atoms in MoS$_2$. The outer S atoms also show little less strong negative EDD than those in MoS2. In the WS$_2$ NT-CNT system also, a tendency to transfer electrons from the W and S atoms, towards the C atoms at the junction is seen. In terms of the change in the EDD of the constituent nanotubes in the combined system, it is observed that at the junction the region of negative EDD within the diameter of the CNT is enhanced due to the presence of the Mo and S atoms. Also the region of positive EDD along the C-C bonds in the CNT side undergoes a shift from the center of the C-C bond, towards the C atom cores at the junction.

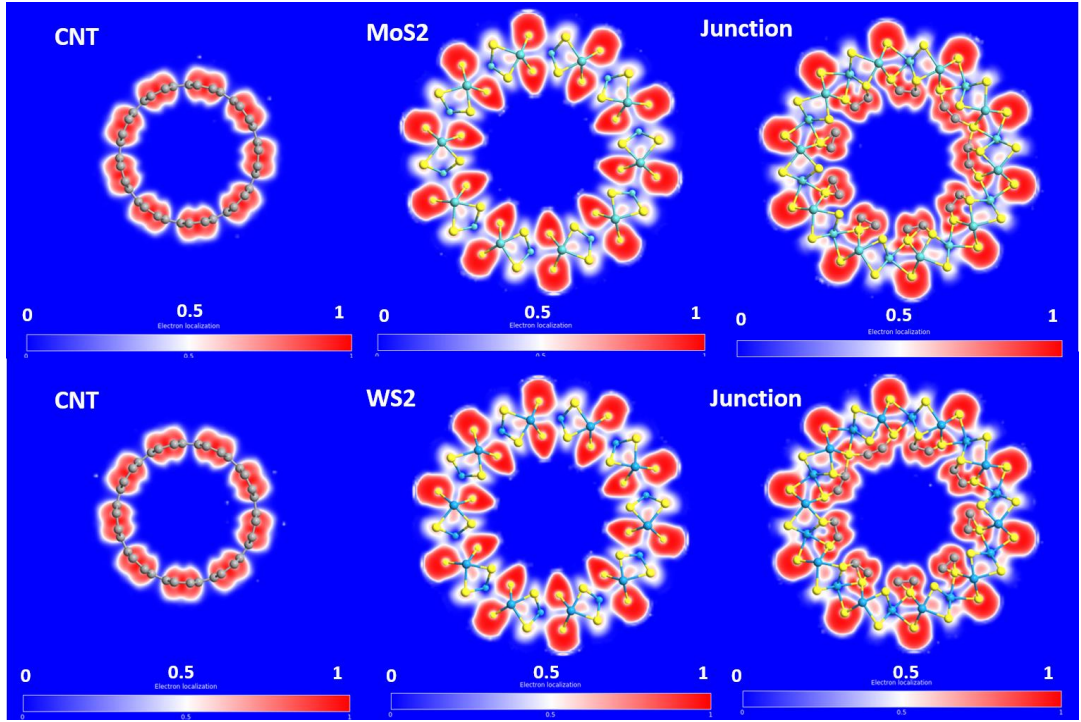

**Fig. 5:** The electron localization function (ELF) plots in the x-y cut plane for the $MX_2$ NT-CNT junctions.

Fig. 5 shows the electron localization function (ELF) [24] of the $MX_2$ NT-CNT junctions presented in the form of plots in the *xy* cut plane. From our calculations on $MoS_2$ NT-CNT junctions it is seen that electrons are highly localized (values of ~1) around the S and C atoms while they are significantly delocalized (values~0) near the Mo atoms. Regions between the Mo-S bonds show presence of an almost uniform electron cloud (~0.5). For the $WS_2$ NT-CNT system, we see a similar pattern of ELF as that of $MoS_2$-CNT but with a stronger delocalization of electron near the W atoms and a sparser presence of electron cloud near the W-S bonds. In both cases at the junction, the electron cloud of S is shifted to cover the C atoms as well.

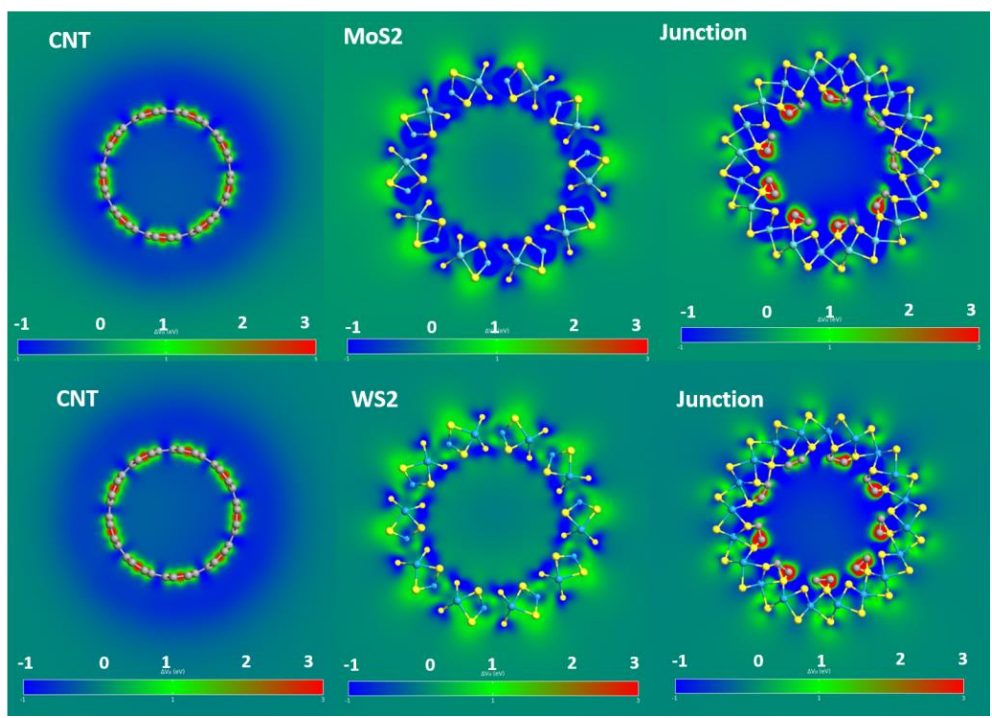

**Fig. 6:** The electrostatic difference potential (EDP) plots in the x-y cut plane for the MX2 NT-CNT junctions.

The electrostatic difference potential (EDP) calculated for the $MX_2$ NT-CNT junctions, are shown in Fig. 6. Strong positive EDP is observed near the C atoms compared to an overall negative EDP around the inner S atoms of the $MoS_2$ NT-CNT junction. These results point to the electron redistribution from the S towards the C atom in the $MX_2$ NT-CNT systems. The in-plane Mo atoms show a negative EDP while the S atoms display slight negativity in an ambience of positive EDP. The $WS_2$ NT-CNT junction shows a different behavior in the sense that here the W atoms lie in a region of slightly more positive EDP. Overall the EDP results suggest the formation of small dipole moment between the chalcogen/ metal and the C atom at the junction interface.

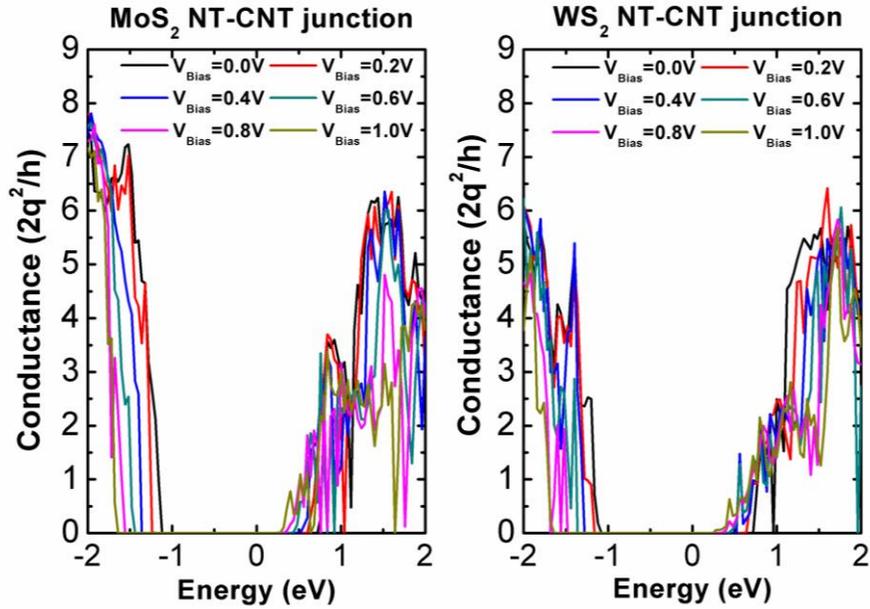

**Fig. 7:** Energy resolved conductance of the MX2 NT-CNT junctions

The energy resolved transmission shown in Fig. 7 displays a marked variation as the system is driven out of equilibrium. The transmission states are shifted from the conduction towards the valence bands as the bias pushes down the quasi-fermi level of the junction. With this shift stronger transmission states in the conduction band come into the bias window while some of the states in the valence band side are pushed out of it. The difference in conductance in the two $MX_2$ NT-CNT junction lies in the larger magnitude of transmission below the Fermi level in $MoS_2$ NT-CNT junction than the $WS_2$ NT- CNT junction. However the no of transmission channels are more spread out in case of the $WS_2$ NT- CNT junction and as we shift to higher voltages, with more and more of the hole transmission states being pushed out of the bias window, the effective current transmission capacity of the $WS_2$ NT- CNT is expected to overrun that of the $MoS_2$ NT-CNT junction.

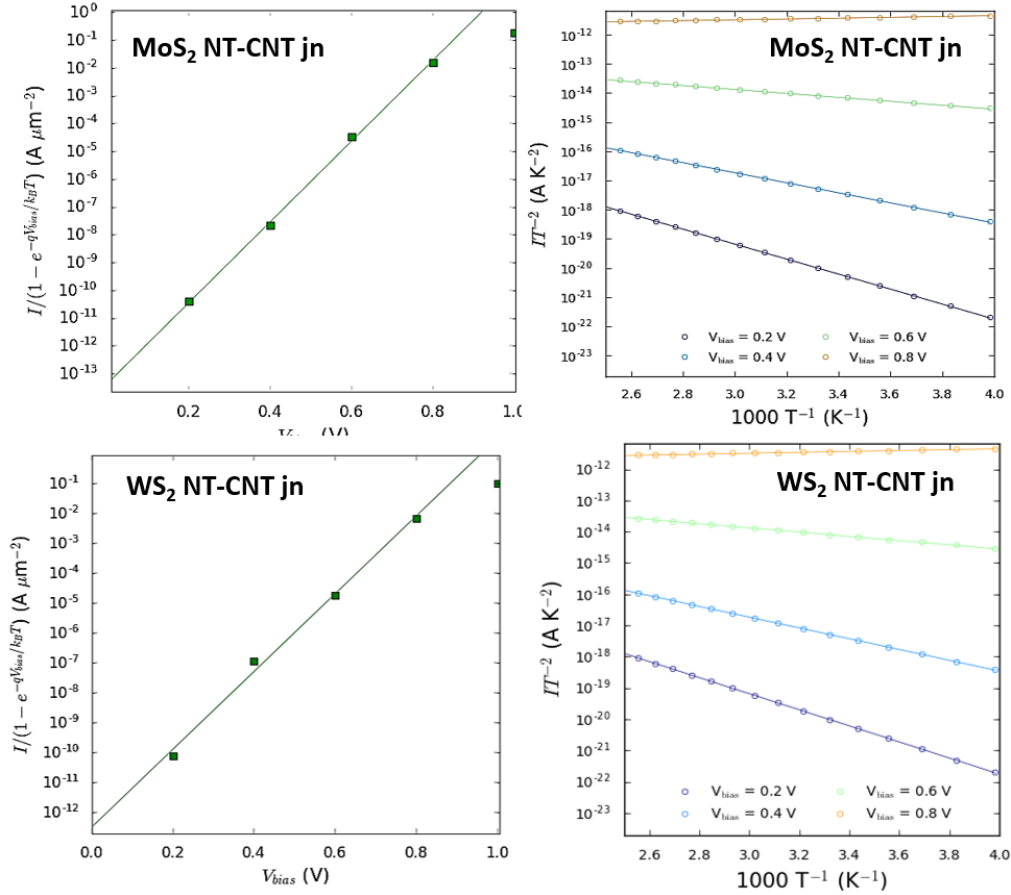

**Fig. 8:** Fitting for the ideality factor extraction and the Arrhenius plots for the calculation of barrier height by activation energy method for the MX2 NT-CNT junctions

In Fig. 8 the fitting data of the equation (1) for extraction of ideality factor $\eta$ is shown alongwith the Arrhenius plots for the estimation of $\Phi^{AE}$ for different bias voltages. The value of $\eta$ is calculated to be 1.1322 for MoS$_2$ NT-CNT junction and 1.2526 for WS$_2$ NT-CNT junction, suggesting more deviation from Schottky diode characteristics for the latter. The current values at biases in excess of 0.8V however do not fit well with this $\eta$ suggesting larger departure from ideal behavior at high applied forward bias. The calculated values of the $\Phi^{AE}$ is listed in Table –I. A notable fact is the bias induced barrier lowering is observed for both the junctions. The computed values of $\Phi^{AE}$ vary from 0.697 to 0.664 eV and 0.669 to 0.610 eV for MoS$_2$ NT-CNT amd WS$_2$ NT-CNT respectively over a bias range of 0.2 to 0.8V. This can be explained by the Schottky barrier inhomogeneity as observed from the PLDOS, showing patches of high LDOS. It is in these patches of high DOS where charge can concentrate and affect the barrier height [19,21]. Another possible cause is the lower density of states of CNT compared to elemental metals, which allows for the easier variation in Fermi level on the CNT side with applied bias [18]-[21]. Also the possibility of quantum tunnelling that is included in the NEGF simulations are not covered in the thermionic emission formula for the Schottky junction, which forms the basis of the parameter extraction in this case. [18]

**Table-I:** The calculated diode parameters for the MX$_2$ NT-CNT junctions.

| Junction | Ideality factor $\eta$ | $RA_{eff}$ ($\Omega.\mu m^2$) | $V_{Bias}$ (V) | $\Phi^{AE}$ (eV) |
|---|---|---|---|---|
| MoS$_2$ NT-CNT | 1.1322 | 7.49x10$^5$ | 0.2 | 0.697 |
|  |  |  | 0.4 | 0.686 |
|  |  |  | 0.6 | 0.678 |
|  |  |  | 0.8 | 0.664 |
| WS$_2$ NT-CNT | 1.2526 | 7.22x10$^5$ | 0.2 | 0.669 |
|  |  |  | 0.4 | 0.661 |
|  |  |  | 0.6 | 0.613 |
|  |  |  | 0.8 | 0.610 |

## 4. CONCLUSION

In this work we have carried out atomistic simulations on the MX$_2$ (MoS$_2$ and WS$_2$) armchair nanotube-CNT junctions. The density functional theory simulations on the (10,10) MX$_2$ NT- (10,10) CNT junctions showed a significant charge accumulation at the junctions with a preference for charge (electron) transfer from metal and sulfur atoms towards the C, which promote dipole moment formation at the junction interface. Also at the junction an enhanced electron localization around the sulfur atoms is observed. An inhomogeneous Schottky barrier formation due to the unique geometry of the interface is evident from the projected local density of states (PLDOS) results. From carrier transport studies based on the DFT-NEGF method Schottky diode parameters were extracted and the MoS$_2$ NT-CNT showed a more ideal junction with WS$_2$ NT-CNT junction having a larger modulation of the barrier with variation in bias. All in all, the results provide interesting insights into the junction physics of this 1D-1D contact, which may be useful in optoelectronics and photonics.

## ACKNOWLEDGEMENT

The work is supported by the Hanse-Wissenschaftskolleg (HWK) fellowship in energy research 2016-17. Part of the work was carried out at Indian Institute of Engineering Science and Technology, Shibpur, with funding from the DST INSPIRE Faculty Grant No. IFA-13 ENG-62.

## ACKNOWLEDGEMENT

The work is supported by the Hanse-Wissenschaftskolleg (HWK) fellowship in energy research 2016-17. Part of the work was carried out at Indian Institute of Engineering Science and Technology, Shibpur, with funding from the DST INSPIRE Faculty Grant No. IFA-13 ENG-62.